\documentclass[amsmath,amssymb,aps,prl,twocolumn,superscriptaddress,showpacs,floatfix]{revtex4-1}

\usepackage{graphicx}

\begin{document} 

\title{Kondo resonance of a Co atom exchange coupled to a ferromagnetic tip}

\author{D.-J. Choi}
\affiliation{IPCMS, CNRS UMR 7504, Universit\'{e} de Strasbourg, 67034 Strasbourg, France}
\affiliation{CIC nanoGUNE, 20018 Donostia-San Sebasti\'{a}n, Spain}
\author{S. Guissart}
\affiliation{Laboratoire de Physique des Solides, CNRS UMR 8502, Universit\'{e} Paris-Sud 11, 91405 Orsay, France}
\author{M. Ormaza}
\author{N. Bachellier}
\author{O. Bengone}
\affiliation{IPCMS, CNRS UMR 7504, Universit\'{e} de Strasbourg, 67034 Strasbourg, France}
\author{P. Simon}
\affiliation{Laboratoire de Physique des Solides, CNRS UMR 8502, Universit\'{e} Paris-Sud 11, 91405 Orsay, France}
\author{L. Limot}
\affiliation{IPCMS, CNRS UMR 7504, Universit\'{e} de Strasbourg, 67034 Strasbourg, France}
\email{limot@ipcms.unistra.fr}

\date{\today}

\begin{abstract}
The Kondo effect of a Co atom on Cu(100) was investigated with a low-temperature scanning tunneling microscope using a monoatomically sharp nickel tip. Upon a tip-Co contact, the differential conductance spectra exhibit a spin-split asymmetric Kondo resonance. The computed \textit{ab initio} value of the exchange coupling is too small to suppress the Kondo effect, but sufficiently large to produce the splitting observed. A quantitative analysis of the line shape using the numerical renormalization group technique indicates that the junction spin polarization is weak. 
\end{abstract}

\maketitle 


The observation of the Kondo effect in quantum dots and single-magnetic impurities (atoms and molecules) placed on a surface or captured between electrodes~\cite{li98,madhavan98,ternes09,scott10} has renewed experimental and theoretical interest in this correlated quantum state. The Kondo effect arises due to conduction electrons scattering off the impurity spin and produces a many-body spin singlet below a characteristic temperature, the Kondo temperature ($T_\text{K}$)~\cite{hewson97}. The most prominent fingerprint of this state is a narrow resonance in the impurity density of states at the Fermi energy. Interestingly, the local magnetic environment of the impurity can alter the ideal line shape of this so-called Kondo resonance. Through a line shape analysis, it is then possible to sense a rich variety of magnetic phenomena at the nanoscale, which include magnetic interactions of the Kondo impurity to surrounding impurities~\cite{heersche06,wahl07,otte09,tsukahara11,bork11,spinelli15,ormaza16,khajetoorians16}, magnetic anisotropy~\cite{otte08,oberg14,khajetoorians15,dubout15,jacobson15} and spin-polarized tunneling electrons~\cite{fu12,bergmann15}.

Kondo and ferromagnetic electron correlations can coexist and compete to influence the ground state of a single-magnetic impurity. When a Kondo impurity is hybridized with a ferromagnetic electrode, the unbalance between spin-up and spin-down states in the host metal should cause the resonance to split into two asymmetric peaks~\cite{lopez03,martinek03,martinek03b}. Introducing ferromagnetism into the Kondo system is, however, difficult in practice, experimental observations of this kind remaining limited and contradictory. A spin-split resonance was successfully observed in quantum dots with ferromagnetic electrodes~\cite{pasupathy04}, or in adsorbates coupled to ferromagnetic nanostructures~\cite{kawahara10,fu12}, while a single Kondo resonance was reported in atomic-scale contacts~\cite{calvo09,neel10}; the asymmetry of the peaks could not systematically be observed. Additional control as well as an improved description of the ferromagnetic environment is therefore desirable for drawing a comprehensive picture. 

Scanning tunneling microscopy (STM) offers the possibility of building a well-defined single-atom contact exhibiting the Kondo effect~\cite{neel07}, which may be tuned through a tip displacement~\cite{choi12}. Here, we use a  nickel tip to contact an individual Co atom adsorbed on a Cu(100) surface (see inset of Fig.~\ref{fig3}), which was recently recognized to be a spin-1/2 Kondo system~\cite{baruselli15,jacob15}. We show that the ferromagnetic exchange coupling between the tip-apex atom and the Co atom inherent to our contact measurement promotes a reproducible asymmetric spin-split Kondo resonance. We carry out a quantitative line shape analysis based on the numerical renormalization group (NRG) technique and extract a spin polarization for the junction. The results are discussed in view of density functional theory (DFT) calculations, which allow estimating the interatomic exchange coupling between the nickel apex atom and the Kondo impurity. A good agreement is found between the two techniques.


\begin{figure}[t]
 \includegraphics[width=0.48\textwidth,clip]{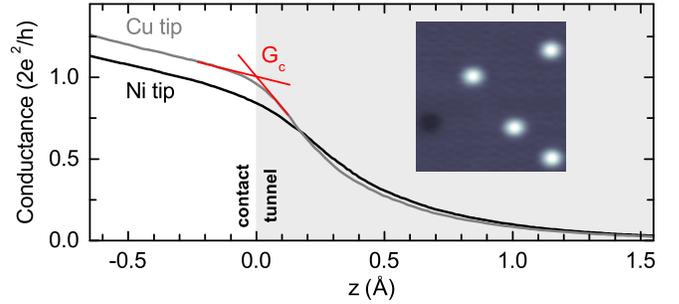}
  \caption{Approach curve above a Co atom recorded with Ni (solid black line) and Cu (solid grey line) tips; the tunneling regime is indicated by a grey background. The curves are acquired at a fixed bias of $V=-160$~mV. To extract the contact conductance, $G_c$, we follow \cite{neel09} and approximate the conductance data in the contact and tunneling regions by straight lines (solid red lines). Their point of intersection defines $G_c$. The inset shows a topographic image of single Co atoms on Cu(100) (100 pA, 100 mV, $8.5\times8.5$~nm$^2$).
\label{fig1}}
\end{figure}

An ultra-high vacuum STM operating at $4.4$~K was used for the measurements. The Cu(100) surface, as well as the tungsten tips, were cleaned \textit{in vacuo} by sputter/anneal cycles. The tungsten tips were further prepared by indentation into the surface to cover their apex with copper (hereafter, we designate them as Cu tips). The nickel tips were only sputter cleaned and subsequently placed close to a Neodymium permanent magnet. Extreme care was taken to maintain the Ni tip apices clean during measurements; chemical control of the apex was routinely ensured through the contact conductance (see Fig.~\ref{fig1}). Single-cobalt atoms were evaporated through openings in the cryostat shields on the cold surface by heating a Co wire (99.99\% purity), which resulted in a coverage of $5\times10^{-3}$ monolayers. 


\begin{figure}
\includegraphics[width=0.48\textwidth,clip]{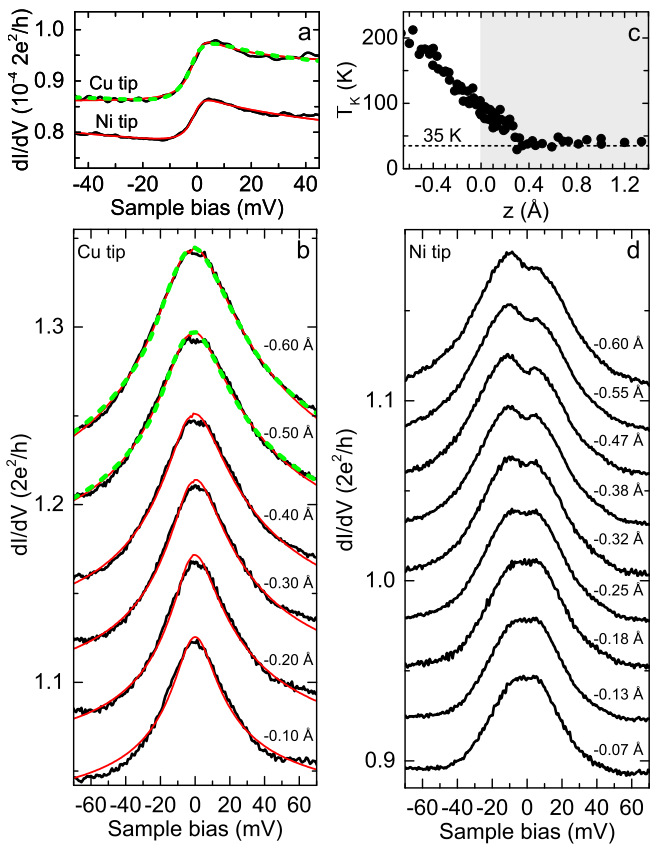}
  \caption{a) $dI/dV$ spectra acquired in the tunneling regime above a Co atom with Ni and Cu tips ($z=4$~{\AA}) and corresponding Frota-Fano fits (solid red lines); the dashed green line is the NRG simulation. The spectrum acquired with the Ni tip is displaced downward by $5\cdot10^{-5}$ in units of $2e^2/h$. b) Set of $dI/dV$ spectra acquired with a Cu tip in the contact regime for different tip excursions (indicated on the right of the panel).  A higher tip excursion produces a higher background in the $dI/dV$. The solid red lines correspond to Frota-Fano fits; the dashed green lines are the NRG simulations. c) $T_\text{K}$ versus $z$ extracted with the Frota-Fano fits. The dashed line corresponds to the Kondo temperature in the tunneling regime.  d) Set of $dI/dV$ spectra acquired with a Ni tip in the contact regime for various tip excursions.
  \label{fig2}}
\end{figure}

Figure~\ref{fig1} presents the typical evolution of the conductance when the tip is vertically displaced towards the center of a Co atom (the tip displacement is noted $z$ hereafter). In the figure, we focus on the transition between the tunneling ($z>0$) and the contact regimes ($z<0$). The average contact conductance is $G_{c}=0.91\pm0.05$ (in units of $2e^2/h$) for pristine Ni tips, while this conductance increases to $G_{c}=1.04\pm0.05$ for Cu tips in agreement with previous studies~\cite{limot05,neel09}. These values demonstrate that the tips employed have a monoatomically sharp apex~\cite{tao10}. The contact geometry corresponds therefore to a bottleneck structure comprising an atom at the tip apex and the Co atom on Cu(100) (see inset of Fig.~\ref{fig3}a for the nickel tip).

The differential conductance ($dI/dV$) versus sample bias ($V$) was measured using a lock-in amplifier (modulation: $500$~$\mu$V~rms, frequency of $712$~Hz) at selected tip excursions above a Co atom. The tip was verified to have a flat electronic structure in the bias range presented. Typical spectra acquired in the tunneling regime ($z=4$~{\AA}) are presented in Fig.~\ref{fig2}a. For both Ni and Cu tips, a single resonance is evidenced near the Fermi level, the step-like shape resulting from the interference between tunneling into the Kondo resonance and tunneling directly into the substrate~\cite{li98,madhavan98,ternes09}. The resonance is well described by a Frota-Fano function (solid red lines in Fig.~\ref{fig2}a)~\cite{frota92,seridonio09,pruser11}; we find $T_\text{K}=(35\pm5)$~K~\cite{n1} and a Fano parameter $q=(1.8\pm0.2)$ for both tips. Hence, when a vacuum barrier is present the Kondo system studied, which is in a strong coupling regime as $T\ll T_\text{K}$, is insensitive to the ferromagnetic nature of the tip. 

In the contact regime ($z\le 0$), the Kondo line shape changes (Fig.~\ref{fig2}b). With Cu tips, a single peak-like resonance is detected, the Frota-Fano fits yielding $q\gg1$. The resonance width also increases monotonically when decreasing the tip-Co distance due to tip-induced modifications of the Co adsorption on Cu(100)~\cite{choi12,baruselli15}; the corresponding Kondo temperatures extracted from the fits are shown in Fig.~\ref{fig2}c and reach values of $200$~K at $z=-0.6$~{\AA} (higher tip excursions can result in tip instabilities). In contrast, with a Ni tip the Kondo resonance splits apart into two peaks (Fig.~\ref{fig2}d). The height of the two peaks differs and changes with tip excursion, resulting in an asymmetric line shape. The peak separation is $\approx15$~mV at the highest tip excursions investigated, all the Ni tips tested producing a similar splitting (see Fig.~\ref{fig4}b). 


A magnetic field is known to break the spin symmetry of a Kondo system and to split apart its resonance. The effective magnetic field produced by the stray field of a Ni tip can be excluded as it is $<0.6$~T~\cite{pasupathy04}, which would correspond to a splitting $<0.2$~mV~\cite{costi00}. Our observations denote instead the existence of an exchange field due to the ferromagnetic interaction between Co and the Ni tip apex. This interaction competes with the antiferromagnetic coupling $J_\text{K}$ between Co and the itinerant electrons of the copper surface, which ensures the Kondo physics~\cite{hewson97}. There are therefore three pertinent energy scales in the system. Two of them, $J_\text{K}$ and $k_\text{B}T_\text{K}$ ($k_\text{B}$: Boltzmann constant), are related to the Kondo physics, and are linked to one another [see Eq.~(\ref{TK})]. The third energy scale is given by the ferromagnetic Ni-Co interaction $J_{\text{fm}}$. We show below that the Kondo effect is preserved as $J_{\text{fm}}\ll\lvert J_\text{K}\rvert$, but $J_{\text{fm}}$ is sufficiently large compared to $k_\text{B}T_\text{K}$ to spin split the Kondo resonance.

\begin{figure}[t]
\includegraphics[width=0.48\textwidth,clip]{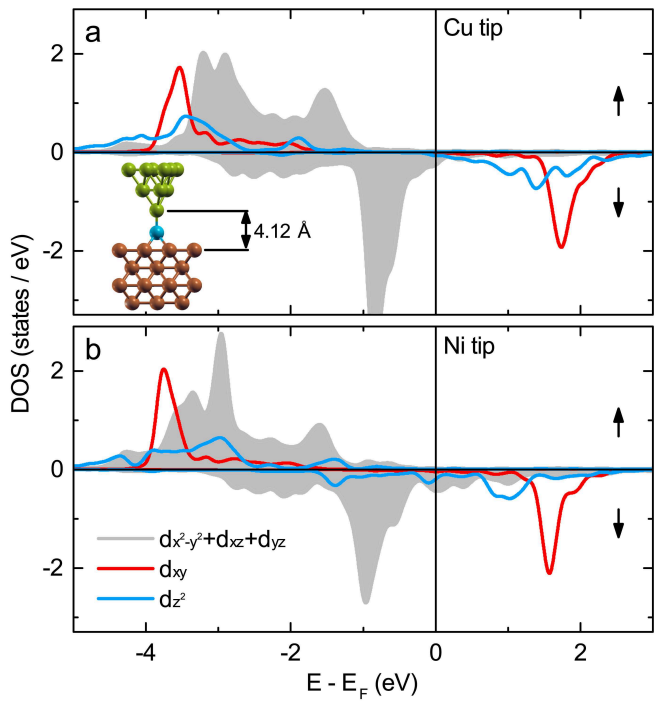}
\caption{$\ell m$-decomposed DOS for a Co/Cu(100) in contact with a) a Cu and b) a Ni tip  (up arrow: majority spins; down arrow: minority spins). The light grey area corresponds to the sum of $d_{yz} + d_{xz}+d_{x^2−y^2}$ doubly occupied orbitals, while the solid red and blue lines correspond to the singly occupied $d_{xy}$ and $d_{z^2}$ orbitals, respectively. Inset of panel a): Geometry used for the DFT calculations. After relaxation, the distance between the apex atom and cobalt is $2.4$~{\AA} with the Cu tip and $2.3$~{\AA} with the Ni tip.
  \label{fig3}}
\end{figure}


To estimate the coupling $J_\text{K}$, we need to take into account the multi-orbital nature of the Kondo effect of the Co atom on copper~\cite{huang08,jacob09,surer12}. With this in mind, we performed DFT calculations for a Cu and a Ni tip in contact with Co/Cu(100). We used the VASP package~\cite{kresse96} within the PAW formalism~\cite{blochl94,kresse99}. To account for the quasi-atomic character of the $3d$ orbitals of Co/Cu(100)\textemdash and describe the corresponding correlation effects, we employed the so-called GGA+U approach~\cite{perdew08,dudarev98,bengone00} with $\overline{U}=3$~eV~\cite{jacob15,frank15}. The geometry used is shown in the inset of Fig.~\ref{fig3}a. The Cu(100) substrate is described by a slab with a $3\times 3$ supercell containing five layers of $9$ atoms in each atomic plane with a lattice parameter of $3.61$~{\AA}. A Co atom is placed above the hollow site of the top substrate layer. The tip is represented by a three-layer pyramid of $10$ atoms arranged in the fcc(111) stacking and terminated by a single apex atom positioned on top of the Co atom~\cite{neel07}. The distance between the tip-apex atom and the first Cu(001) layer is fixed to $4.12$~{\AA} (see the double-headed arrow in the inset of Fig.~\ref{fig3}a), which is representative of the contact regime~\cite{baruselli15}. All other atoms of the cell are allowed to relax.

Figure~\ref{fig3}a presents the $\ell m$-decomposed density of states (DOS) projected on the Co atom for the case of the Cu tip. The DOS associated to the $d_{z^2}$ and $d_{xy}$ orbitals is clearly spin polarized, giving a magnetic character to the cobalt atom, while the rest of the $d$ shell is occupied and nonmagnetic. We find two singly occupied $d_{z^2}$ and $d_{xy}$~\cite{n5} orbitals, in agreement with earlier calculations on this system~\cite{huang08,baruselli15,jacob15,supp}, resulting in a total spin of $S\approx1$ for Co/Cu(100). Both orbitals are susceptible to produce a Kondo effect. We find, however, that the experimental data are well described by the numerically exact spin-$1/2$ line shape computed with the NRG method~\cite{supp}, in both tunneling and contact regimes (dashed lines, respectively, in Figs.~\ref{fig2}a and~\ref{fig2}b). This indicates that at our working temperature only one of the two orbitals is Kondo screened. Recent calculations suggest that the $d_{z^2}$ orbital is most likely responsible for the Kondo effect~\cite{baruselli15,jacob15,n2}, while the spin in the $d_{xy}$ orbital remains unscreened at accessible temperatures. The coupling $J_\text{K}$ can then be extracted via the experimental Kondo temperature by recalling that for a spin-1/2 Kondo system we have~\cite{hewson97} 
\begin{equation} 
k_\text{B}T_\text{K}=W\sqrt{2\rho \lvert
J_\text{K}\rvert}\exp{\left(\frac{1}{2\rho J_\text{K}}\right)}, 
\label{TK} 
\end{equation}
where $\rho=0.3$~eV$^{-1}$ is the density of states of copper at the Fermi energy; we follow \cite{ujsaghy00} and use as a band cutoff $W\simeq \epsilon^\text{Cu}_\text{F}=7$~eV. For $T_\text{K}=200$~K, which corresponds to $z=-0.6$~{\AA} (Fig.~\ref{fig2}c), we find $J_\text{K}=-0.32$~eV.

The GGA+U calculations with the Ni tip yield a similar $\ell m$-decomposed DOS (Fig.~\ref{fig3}b) and, in particular, the same $d$-orbital occupations as with the Cu tip or other tip geometries~\cite{supp}. The STM data indicates that the Kondo line shape changes in the presence of a Ni tip, but, as we show below, it is still well described by a spin-1/2 model. This again suggests that only one of the two $d$ orbitals is responsible for the the Kondo effect observed. Given the similarity for the Cu- and Ni-tip calculations, in the following we will suppose that the coupling $J_\text{K}$ determined is representative for both Kondo systems.

The next step of our analysis consists in computing the ferromagnetic exchange coupling $J_{\text{fm}}$ between the Co atom and a neighboring Ni atom. To do so, we performed additional DFT calculations in the framework of the tight-binding linear Muffin-Tin orbital (TB-LMTO) method generalized to surfaces and interfaces~\cite{turekbook97}, using the surface Green function formalism. The TB-LMTO method allows evaluating directly the magnetic exchange interaction between atoms in the system~\cite{liechtenstein87,pajda01,ondracek10}, at the expense of a rough approximation of the junction geometry. The system was in fact modeled with a $3\times3$ lateral supercell with the Co atom sandwiched between the (100) facet of a Cu and of a Ni fcc crystal of lattice parameter $3.61$~{\AA}. The cobalt atom is placed at the hollow site of   both surfaces at a distance of $1.81$~{\AA} from the Ni/Cu planes. In this geometry the Ni tip may be considered as blunt. Interestingly, we find a ferromagnetic coupling of $J_{\text{fm}}=15.3$~meV between a nickel and a cobalt atom (Ni-Co distance: $2.55$~{\AA}). This value is several orders of magnitude smaller than $J_\text{K}$, thereby confirming our experimental findings that the Kondo effect of Co is preserved upon contact with the Ni tip.

The Kondo resonance can however be expected to split apart in view of the exchange field ($B_\text{ex}$) associated to $J_{\text{fm}}$. Within mean-field theory, we have $B_\text{ex}=n J_{\text{fm}} \langle S_\text{tip}\rangle/g\mu_\text{B}$, where $\langle S_\text{tip}\rangle=0.6$~$\mu_\text{B}$ is the magnetic moment of the Ni atom; we take $n=1$ as the coordination number since the Co atom in the experimental contact geometry has only one neighboring nickel atom. Assuming a $g$ factor of $2$, the exchange field amounts to $79$~T and would yield a splitting of $e\Delta V = 2g\mu_\text{B}B_\text{ex}\simeq 18$~meV~\cite{costi00}, close to experimental findings. We anticipate that a similar splitting can be expected with other ferromagnetic tips as the value of $J_{\text{fm}}$ found here is typical for common ferromagnets~\cite{pajda01}. We also note that $J_{\text{fm}}$ weakens with increasing Ni-Co distance, a separation of $3.61$~{\AA} resulting in $J_{\text{fm}}=1.5$~meV. The splitting in this case decreases beyond detectability, in agreement with the single Kondo resonance that is detected in the tunneling regime at $z=4$~{\AA} (Fig.~\ref{fig2}a). This does not exclude probing an exchange field in the tunneling regime with Kondo systems possessing a narrower resonance than the present one.

\begin{figure}[t]
\includegraphics[width=0.48\textwidth,clip]{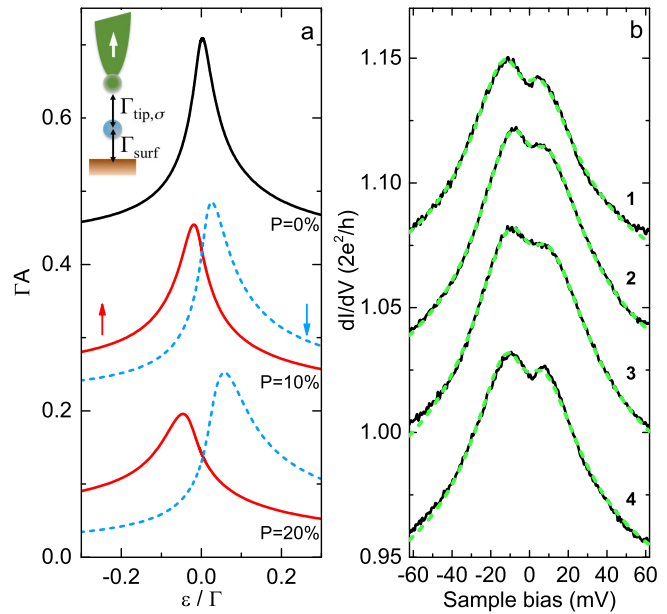}
\caption{a)
Functions $A_\uparrow$ (solid red line) and $A_\downarrow$ (dashed blue line) for several $P$ ($A_\uparrow=A_\downarrow$ for $P=0\%$). The curves are shifted vertically for clarity ($P=0\%$ is shifted by 0.40 and $P=10\%$ by 0.2). Inset: Schematic view of the model employed. b) $dI/dV$ spectra acquired with four different Ni tips (labeled \textbf{1} to \textbf{4}) and corresponding NRG fit using Eq.~(\ref{nrg}) (solid red lines). For clarity, the spectra \textbf{2}, \textbf{3} and \textbf{4} are shifted vertically by 0.04, 0.05 and 0.1 respectively. \label{fig4}}
\end{figure}


To confirm the presence of an exchange interaction when using a Ni tip, we use the NRG method to numerically simulate the exact resonance line shape. We use a spin-1/2 model where
the presence of ferromagnetism is accounted for by a spin-dependent hybridization~\cite{lopez03,martinek03,martinek03b,patton07,seridonio09,lim13}; a full description of the Hamiltonian is given as Supplemental Material~\cite{supp}. Within the paradigm proposed, the spin-dependent hybridization is carried by the ferromagnetic tip through the hybridization function $\Gamma_{\text{tip}}=(\Gamma_{\text{tip},\uparrow}+\Gamma_{\text{tip},\downarrow})/2$ (inset of Fig.~\ref{fig4}a), while the hybridization function of the surface ($\Gamma_\text{surf}$) is unpolarized (the total hybridization is $\Gamma=\Gamma_{\text{tip}}+\Gamma_{\text{surf}}$). We then compute the differential conductance by using Wilson's NRG method and fit the experimental data with~\cite{zitko11,n3} 
\begin{equation}
g(V)=\sum_{\sigma=\uparrow,\downarrow}h_\sigma A_\sigma(eV)+g_0, 
\label{nrg}
\end{equation}
where $g_0$ is a constant background (the expression is given in units of $2e^2/h$). The simulated spectrum $g(V)$ is the sum of two fully spin-polarized functions $A_\uparrow$ and $A_\downarrow$~\cite{martinek03b}, each weighted by a spin-dependent prefactor, respectively, $h_{\uparrow}$ and $h_{\downarrow}$. As shown in Fig.~\ref{fig4}a, $A_\uparrow$ and $A_\downarrow$ have an asymmetric line shape and are displaced, respectively,  downward and upward relative to the Fermi level, their separation increasing with the junction polarization $P=(\Gamma_{\text{tip},\uparrow}-\Gamma_{\text{tip},\downarrow})/\Gamma$. The fit to the data based on Eq.~(\ref{nrg}) is presented in Fig.~\ref{fig4}b and is highly satisfactory (see Supplemental Material for details on the fitting procedure). The experimental spectra of Fig.~\ref{fig4}b were recorded in the contact regime ($z=-0.6$~{\AA}) using different Ni tips. We find an average junction polarization $P=(7\pm1)\%$, while $\Gamma=0.23$~eV; this results in a splitting $e\Delta V\simeq P\Gamma$~\cite{martinek03,patton07,seridonio09,supp} of $(16\pm2)$~meV. The extracted values for the prefactors yield instead a spin asymmetry of $\eta=(h_{\uparrow}-h_{\downarrow})/(h_{\uparrow}+h_{\downarrow})=(22\pm2)\%$. We note that both $P$ and $\eta$ reflect a spin polarization, but are impacted by the impurity hybridization to the tip and the surface in a different way~\cite{n4}.


To summarize, we have shown that the Kondo effect in a single-atom contact subsists in the presence of a ferromagnetic tip, but the Kondo line shape is spin split. Our findings support the assignment of a Kondo effect in bulk ferromagnetic constrictions~\cite{calvo09,neel10}, but possibly indicate that the Kondo resonance may be spin split in these systems. We stress that the setup employed here can be easily generalized to other atoms or molecules in order to explore the interplay between the  Kondo effect and ferromagnetism, and eventually other surface-supported magnetic effects.

\begin{acknowledgments}
We thank M.V. Rastei, F.\ Scheurer, N.\ Lorente and R. \u{Z}itko for fruitful discussions. This work was supported by the Agence Nationale de la Recherche through Grant No. ANR-13-BS10-0016, ANR-11-LABX-0058 NIE, ANR-10-LABX-0026 CSC. This work was performed using the HPC resources of the Meso Center of the University of Strasbourg.
\end{acknowledgments}

\bibliography{Kondo_FM}

\begin{thebibliography}{62}
\bibitem [{n1()}]{n1}%
  \BibitemOpen
  \href@noop {} {}\bibinfo {note} {The Kondo temperature has various
  definitions in literature. Here, the intrinsic width of the Frota line
  corresponds to $1.455 k_\text{B}T_\text{K}$~\cite{pruser11}. We find Kondo
  temperatures $2.5$ times lower than previous
  studies~\cite{wahl07,neel07,choi12} where Lorentzian-based fits were
  used.}\BibitemShut {Stop}%
\bibitem [{n5()}]{n5}%
  \BibitemOpen
  \href@noop {} {}\bibinfo {note} {The $xy$-axis was chosen along the second
  nearest-neighbors direction of Cu(100). If the $xy$-axis points along the
  first nearest-neighbors direction, $d_{xy}$ becomes $d_{x^2-y^2}$ as
  in~\cite{huang08}.}\BibitemShut {Stop}%
\bibitem [{n2()}]{n2}%
  \BibitemOpen
  \href@noop {} {}\bibinfo {note} {Magnetic anisotropy, which is $\lesssim1$
  meV for $3d$ atoms on metals~\cite{dubout15,khajetoorians15}, can be
  neglected here in view of the large value of
  $k_\text{B}T_\text{K}$.}\BibitemShut {Stop}%
\bibitem [{n3()}]{n3}%
  \BibitemOpen
  \href@noop {} {}\bibinfo {note} {Equation~(\ref{nrg}) can be further
  generalized to include a quantum interference between a narrow and a broad
  scattering channel, similar to the Fano formula~\cite{supp}. Here we
  exclusively focus onto the contact regime where $q\gg1$.}\BibitemShut {Stop}%
\bibitem [{n4()}]{n4}%
  \BibitemOpen
  \href@noop {} {}\bibinfo {note} {The prefactors of Eq.~(\ref{nrg}) are
  defined as $h_{\sigma}=2\Gamma_{\text{tip},\sigma}/(1+\lambda)$ with
  $\lambda=\Gamma_{\text{tip},\sigma}/\Gamma_\text{surf}$~\cite{supp}. $P$ and
  $\eta$ are therefore both function of $\Gamma_{\text{tip},\sigma}$ and
  $\Gamma_\text{surf}$, which depend on $z$. Note that in the tunneling regime
  where $\lambda\ll1$, we have $h_{\sigma}\simeq2\Gamma_{\text{tip},\sigma}$
  and $\eta$ corresponds to the spin polarization of the tip as in
  \cite{bergmann15}.}\BibitemShut {Stop}%
\end{thebibliography}

\end{document}